\documentclass[12pt]{article}
\usepackage{latexsym}
\usepackage{epsfig,graphics}
\usepackage{graphicx}

\newcommand{\be}{\begin{equation}}
\newcommand{\ee}{\end{equation}}
\newcommand{\ba}{\begin{eqnarray}}
\newcommand{\ea}{\end{eqnarray}}

\newcommand{\eqn}{Eq.~}

\newcommand{\la}{\label}

\begin{document}

\begin{titlepage}

\vspace*{0.7in}

\begin{center}
{\large\bf Glueball Regge trajectories
  and the Pomeron \\
  -- a lattice study -- \\}
\vspace*{1.0in}
{Harvey B. Meyer and Michael J. Teper\\
\vspace*{.2in}
Rudolf Peierls Centre for Theoretical Physics, \\
University of Oxford,\\
1 Keble Road, Oxford OX1 3NP, U.K.\\
}
\end{center}

\vspace*{0.6in}

\begin{center}
{\bf Abstract}
\end{center}

We perform lattice calculations of the lightest
$J=0,2,4,6$ glueball masses in the D=3+1 SU(3) gauge 
theory and extrapolate to the continuum limit.
Assuming that these masses lie on linear Regge 
trajectories we find a leading glueball trajectory
$\alpha(t)=0.93(24) + 0.28(2)\alpha^\prime_R t$, where
$\alpha^\prime_R \simeq 0.9 \, \mathrm{GeV}^{-2}$ is the 
slope of the usual mesonic
Regge trajectory. This glueball trajectory has an intercept
and slope similar to that of the Pomeron trajectory.
We contrast this with the situation
in D=2+1 where the leading glueball Regge trajectory
is found to have too small an intercept to be important
for high-energy cross-sections.
We interpret the observed states and trajectories in
terms of open and closed string models of glueballs.
We discuss the large-$N$
limit and  perform an SU(8) calculation that hints
at new states based on closed strings in higher representations.

\end{titlepage}

\setcounter{page}{1}
\newpage
\pagestyle{plain}

%
%
\section{Introduction}
\label{section_intro}
%
%

The experimentally observed mesons and baryons appear to lie on 
nearly linear  and parallel Regge trajectories, 
\be
J \equiv \alpha(t=m^2)
\simeq \alpha_0 + \alpha^{\prime} m^2 
\label{eqn_regge}
\ee
with $\alpha^{\prime} \simeq 0.9 \, \mathrm{GeV}^{-2}$ and $\alpha_0 \leq 0.5$. 
The exchange of the highest-lying Regge pole will dominate any high energy
scattering that involves the exchange of the corresponding quantum 
numbers (see e.g.~\cite{kaidalov} for a recent review). 

The total cross-section, on the other hand, is related 
by unitarity to forward elastic scattering and this is dominated 
by the `Pomeron' which carries vacuum 
quantum numbers
\cite{kaidalov,book,landshoff}. 
The Pomeron trajectory is qualitatively different from other 
Regge trajectories in that it is
much flatter ($\alpha^{\prime}$ much smaller)
and has a higher intercept
\cite{landshoff}
\be
\alpha_P(t=m^2)
\simeq 
1.08 + 0.25 m^2
\label{eqn_pomeron}
\ee
(in GeV units). A unit intercept would lead to total
cross-sections that are constant with energy. The fact that
cross-sections increase slowly with energy suggests an intercept
slightly larger than unity. Since it does not seem possible 
to associate the Pomeron with the usual flavour-singlet mesons
(whose leading trajectory would have the usual slope and too low
an intercept) there has been a long-standing speculation that
the physical particles on the trajectory (at integer values of the
spin $J$) might be glueballs. This picture arises naturally in
string models of hadrons.

If we now consider the
high-energy scattering of glueballs in the pure SU(3) gauge theory,
it is difficult to imagine that the total cross-section should
behave differently from total cross-sections in the real world.
For instance, in leading-logarithmic perturbative
 calculations (\cite{book} and ref. therein),
only the gluonic field contributes to the Pomeron.
Thus it is reasonable to expect that the Pomeron will appear 
in the pure gauge theory, with similar properties to those
of the phenomenological Pomeron (up to corrections due to 
effects such as mixing). 
This constitutes the main motivation for the calculations of this
paper in which we use numerical lattice techniques to 
investigate whether the mass spectrum of the SU(3) gauge theory 
is consistent with approximately straight Regge trajectories,
the leading one of which possesses the properties of the 
phenomenological Pomeron.

The states that lie on the  phenomenological Pomeron will have even 
spin (the trajectory has even signature) and will start with $J=2$
since the high intercept implies that $m^2 < 0$ for $J=0$ so that
the lightest  $J=0$ state must lie on 
a daughter trajectory. Thus we need to calculate the lightest masses
with $J=2$ and $J=4$, and preferably $J=6$ as well. There are two
major obstacles to this. The first arises from the reduced rotational
invariance of the cubic lattice, which makes the identification of
states with $J\geq 4$ a non-trivial problem. In
\cite{hspin}, 
we developed a technique  to label highly excited  
states from the lattice with the correct spin $J$
and we applied it in the simpler context of 2+1 
dimensional $SU(N)$ gauge theories~\cite{regge_2d}. We have now 
extended this technique to three space dimensions and will use
it in this paper. The technical details are left to a longer
publication
\cite{thesis}.
The second obstacle is that the higher spin
states are much more massive and therefore difficult to
calculate accurately by the standard numerical methods.
We therefore apply recent algorithmic improvements~\cite{2leva, 2levb} 
that help reduce the variance of rapidly decaying correlators. 

In the next Section we summarise the results of our lattice
calculation of the $PC$=++ sector of the glueball spectrum 
and identify the leading and sub-leading glueball Regge 
trajectories. We find that the former does indeed possess
the qualitative features of the Pomeron. To show 
how things might have been different, we also summarise
the results of a similar calculation in $D=2+1$ where one finds a 
leading glueball trajectory that has a very low intercept. We then
turn to a discussion of the string picture of mesons and
glueballs, which provides the framework within which we
interpret our results for the glueball mass spectrum.
(This is what one would now refer to as the `old' string picture 
for QCD; for relevant work within the `new' string picture, see
\cite{Zayas}
and references therein.)
We first remind the reader how the usual mesonic
Regge trajectories can be understood within a simple string
model of  quarkonia, and how this picture naturally translates
to glueballs in the pure gauge theory. We emphasise the richer
structure that this predicts for the associated glueball Regge
trajectories, and we use the observed pattern of states
and degeneracies to associate the observed  trajectories
with specific kinds of open and closed strings.   
As well as discussing the well-established Pomeron trajectory,
we use our calculated spectrum to comment upon the 
more speculative $C$=$-$ odderon (for a review see~\cite{odderon}).
Finally we comment upon the SU($N\to\infty$) limit and some 
implications for the high energy scattering of glueballs and hadrons.

This paper is a summary of the results of calculations that will be 
described in detail in a longer paper
\cite{thesis}. 
In particular the reader is
referred to that paper for the technical details of our lattice 
calculations as well as for a more detailed exploration of
what string models predict for glueballs and for comments
on earlier lattice calculations of higher spin glueballs.

%
%
\section{Results for the $PC=++$ glueball spectrum.}
\label{section_results} 
%
%

Our lattice calculations employ the standard plaquette action.
We calculate ground and excited state masses, $m$, from 
Euclidean correlation 
functions using standard variational techniques. We calculate the
string tension, $\sigma$, by calculating the mass of a flux loop 
that closes around a spatial torus. We perform calculations
for values of the inverse bare coupling $\beta = 6/g^2$
ranging from $\beta=6.0$ to $\beta=6.4$, which corresponds to lattice 
spacings $a \simeq 0.10 - 0.05 \, \mathrm{fm}$. The calculations are on
lattices ranging from $16^3 36$ to $32^3 48$, corresponding to
a spatial extent $aL \simeq 1.5 \, \mathrm{fm}$. At one value of $\beta$ 
we perform calculations on lattices up to $2 \, \mathrm{fm}$ across so as
to check that any finite volume corrections are small. We
extrapolate the calculated values of the
dimensionless ratio $m/\surd\sigma$ to
$a=0$ using an $a^2\sigma$ correction, which is the leading correction with
the plaquette action. We thus obtain the continuum glueball
spectrum with masses expressed in units of the string tension.
All this is quite standard (see e.g. 
\cite{lucini}).

There are two novel aspects to our calculations. The first is
 a recently developed variance reduction technique
\cite{2leva,2levb} 
that is very useful for reducing statistical errors on masses
that are large, such as those of the higher spin states in which 
we shall be interested. The second is the identification of
the lightest $J\geq 4$ states. The problem is that the cubic 
rotation group of the lattice is much smaller than the continuum
rotation group and has just a few irreducible representations. 
Nonetheless this does not mean that it makes no sense to
label states by their `spin $J$'. As $a\to 0$ an energy 
eigenstate belonging to one of these lattice representations
will tend to some state that is labelled by spin $J$. So using 
continuity we can refer to a state at finite $a$ as being of `spin
$J$' if $a$ is small enough. (Level crossings at large $a$ may
eventually render such a labelling ambiguous.) At $a=0$ 
a state of spin $J$ will appear in a multiplet of $2J+1$ degenerate 
states. If we now increase $a$ from zero, these  $2J+1$ states will
in general appear in different lattice representations, and the 
degeneracy will be broken at $O(a^2)$. So in general the ground state of 
spin $J=4,5, 6, ...$ will be a (highly) excited state in some lattice
representation, thus complicating its identification. 
If we can perform this identification, then we can  extrapolate 
the mass of the state to $a=0$, so obtaining the mass of, say, 
the lightest state of spin $J$. Our identification technique, as 
described in
\cite{hspin}
for the simpler case of $D=2+1$, is to perform a Fourier transform
of the rotational properties of any given lattice eigenstate,
using as a probe a set of lattice operators that have an approximate
rotational symmetry that is greater than the exact cubic symmetry,
so that we can probe rotational properties 
under rotations finer than $\pi/2$. If we find that the state
has predominantly the rotational properties
 corresponding to say  $J=4$,
and if we find that this predominance grows towards unity
as $a\to 0$, then we can assign to it these continuum
rotational quantum numbers. In addition there should be
states corresponding to the other members of the spin multiplet
that become degenerate with it as $a\to 0$, and
this provides extra evidence for the correctness of the spin 
assignment. (The density of states and the errors on masses 
prevents us from using this degeneracy as the sole criterion
in practice.) 

The technical details of these methods will be provided elsewhere
\cite{thesis}.
We now turn to a summary of our results.

\subsection{Glueball Regge trajectories in D=3+1}
\label{subsection_4D}
%
%

\begin{figure*}[t]
\vspace{-0.5cm}
\centerline{\begin{minipage}[c]{14cm}
\psfig{file=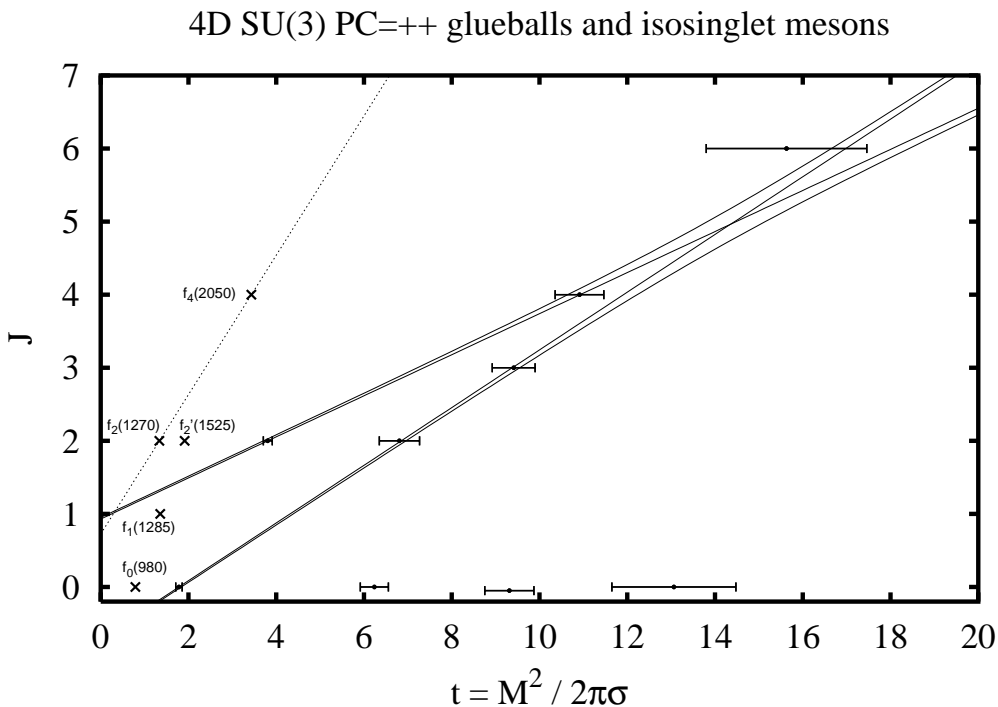,angle=0,width=14cm}	\end{minipage}}
\caption{Chew-Frautschi plot of the continuum 4D $SU(3)$ gauge theory.
The hyperbolae are drawn to suggest the behaviour of the two leading
trajectories. The position of some experimental flavour-singlet mesons is 
indicated~\cite{Hagiwara:fs}.}
\la{fig:cf_pp}
\end{figure*}
\begin{figure*}[t]
\vspace{-0.5cm}
\centerline{\begin{minipage}[c]{14cm}
\psfig{file=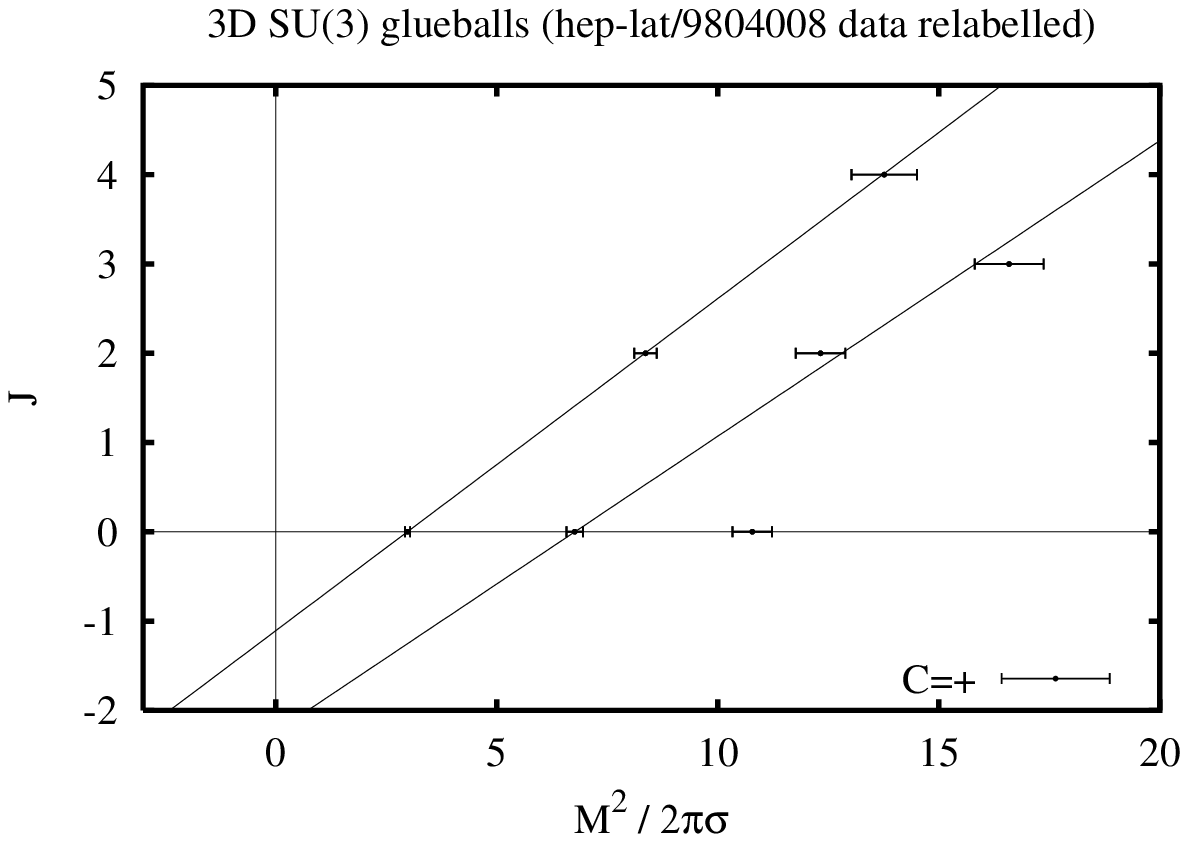,angle=0,width=14cm}	\end{minipage}}
\caption{The Chew-Frautschi plot of the continuum D=2+1 $SU(3)$ 
glueball spectrum.}
\la{fig:cf_su3_3d_pp}
\end{figure*}

We initially focus on states with $PC$=++ since these are the
quantum numbers carried by the Pomeron.

Extrapolating our glueball masses to the continuum limit
we plot the (squared) masses against the spins in a Chew-Frautschi 
plot, as in Fig.1. We now assume that the states fall on 
approximately linear Regge trajectories. (To obtain significant evidence 
for or against such linearity would require more accurate $J=6$ masses 
than we have been able to achieve in the present calculation.)
In that case the leading trajectory clearly passes through the lightest
$J=2$ and $J=4$ glueballs (and within about one standard deviation 
of the lightest $J=6$ glueball). We note that there is no odd $J$ 
state on this trajectory: it is even signature just like the
phenomenological Pomeron. The parameters of the trajectory are 
\be
2\pi\sigma\alpha'=0.281(22)\qquad\alpha_0=0.93(24).
\label{eqn_leadGR}
\ee
which is entirely consistent with the phenomenological Pomeron
in \eqn(\ref{eqn_pomeron}), if we recall that
the usual mesonic trajectories have slopes
\be
\alpha'_{\cal R} \simeq \frac{1}{2\pi\sigma}\simeq 0.9 \, \mathrm{GeV}^{-2}.
\label{eqn_slopeR}
\ee

Of course, in comparing our leading pure-glue trajectory with the 
phenomenological Pomeron we should not ignore the fact that the
latter will mix with the flavourless mesonic trajectory, shown
in Fig.1. It is plausible that the mixing will effectively
increase the intercept and the slope of the Pomeron. In particular
it might well be that the underlying unmixed pure-gauge Pomeron
has an intercept of 1 rather than $\sim 1.08$.

We can also identify in Fig.1 the sub-leading glueball trajectory.
It contains
the lightest $J=0$ glueball, the first excited $J=2$ glueball
and the lightest $J=3$ glueball. Our lower bound on the mass of
the lightest $J^{PC}=1^{++}$ makes it clear that it is much too
heavy to lie on this trajectory. We remark that we have not identified 
any excited $J=4$  or $J=5$ states, with  $PC=++$, and so cannot
say whether they lie on this trajectory or not. In striking
contrast to what one finds for the usual mesonic trajectories,
this secondary trajectory is clearly not parallel to the leading one.
As we shall see in the next Section, this is something one might
expect within a string picture of glueballs. The trajectories cross
somewhere near $J=5$ and it is not quite clear to which trajectory 
the observed $J=6$ state belongs. Clearly it would be useful
to have a mass estimate for the first excited $J=6$ state. Finally
we remark that because of unitarity the trajectories will not
actually cross but will rather repel, as indicated in Fig. 1.

We thus conclude that the leading Regge trajectory in the pure SU(3) 
gauge theory does indeed appear to be the `bare'  Pomeron,
which will become the phenomenological Pomeron after
mixing with the appropriate mesonic Regge trajectory.
We now turn to a similar analysis for the SU(3) 
gauge theory in $D=2+1$, which will demonstrate that
there is nothing inevitable about our above result.

\subsection{Glueball Regge trajectories in D=2+1}
\label{subsection_3D}

In Fig.2 we show the Chew-Frautschi plot for the $C=+$
sector of the continuum SU(3) gauge theory in $D=2+1$.
(We do not refer to parity, because in two space dimensions
one has automatic parity-doubling for $J\not= 0$.)
In contrast to $D=3+1$, a linear trajectory between the 
lightest $J=2$ and $J=4$ states passes through the lightest
$J=0$ state, and so we  should place the $J=0$ 
glueball on that trajectory. Between them the $J=0,2,4$
states provide strong evidence for the approximate linearity
of the trajectory. In contrast to $D=3+1$ the secondary
trajectory is approximately parallel to the leading one. 

It is clear from Fig.2 that the intercept is very low,
so that the leading glueball trajectory will make a
contribution to the total cross-section that decreases
rapidly with energy. Thus if the glueball-glueball
total cross-section is approximately constant at high 
energies, then it will have to be understood
in terms of something other than a Regge trajectory. 

The parameters of the leading trajectory are
\be
2\pi\sigma\alpha'=0.384(16)\qquad\alpha_0=-1.144(71).
\label{eqn_leadGRd3}
\ee
Thus, in contrast to the intercept, the slope of the trajectory
is not very different from what we found in $D=3+1$.

%
%
\section{String models}
\label{section_strings}
%
%

The string picture of hadrons starts from the observation
that linear confinement implies that the flux between
static charges is contained within a flux tube
whose width will be on the order of $1/\surd\sigma$.
Once this flux tube is long enough, 
$l\gg 1/\surd\sigma$, it will look like a string
and this has led to a long-standing conjecture that
the long-distance properties of QCD are given by some
effective string theory. 

The fact that mesons fall on nearly 
linear and parallel Regge trajectories reinforces this picture.
A natural model for a high $J$ meson is to see it as
a rotating string with a $q$ and $\bar{q}$ at its ends
and, as is well-known (see e.g. 
\cite{perkins}),
a simple classical calculation shows how this leads to linear 
Regge trajectories with a slope determined by the string
tension $\sigma$ (the rest energy 
per unit length of the string), 
\be
J \stackrel{J\to\infty}{=} \frac{1}{2\pi\sigma} m^2 + ...
\label{eqn_mesonR}
\ee
The value of the confining string tension that this requires,
in order to produce the observed Regge slope,
is consistent with the string tension one needs for the linear 
part of the heavy-quark potential and with the value that has
been obtained in numerical simulations of quenched and full QCD
(when expressed, for example, in units of the calculated
$\rho$ meson mass or the nucleon mass).

If we now go to the pure gauge theory, this simple `open string'
model has an immediate analogue; two gluons joined
by a string containing flux in the adjoint representation.
However, in contrast to the case of mesons, there is an 
alternative closed string model: a closed string of flux
in the fundamental representation. We now discuss and contrast
these two models in more detail.

\subsection{Open strings}
\label{subsection_open}

We would expect a state of high spin to be highly extended,
and in a confining theory this immediately suggests an open string.
For mesons the string ends on quarks and carries fundamental flux,
while for glueballs it ends on gluons and carries adjoint flux.
Such an adjoint string can break through gluons
popping out of the vacuum, but this is also true of the mesonic
string (through $q{\bar q}$ popping out of the vacuum). What is 
important for the model to make sense is that the decay width should
be sufficiently small -- essentially that the lifetime of the adjoint
string should be much longer than the period of rotation.
In $SU(N)$ gauge theories, both the
adjoint and fundamental strings become completely stable as
$N\to\infty$. So if we are close to that limit the model can
make sense. Indeed adjoint string breaking in $SU(N)$ occurs at
$O(1/N^2)$ while fundamental string breaking in $QCD_{N}$ occurs
at $O(1/N)$. Thus {\it a priori} the
adjoint string model can be taken at least as seriously as the
conventional mesonic string model.

By the same classical calculation as used for mesons, the rotating
adjoint string will produce a linear Regge trajectory
\be
J \stackrel{J\to\infty}{=} \frac{1}{2\pi\sigma_A} m^2 + ...
\simeq \frac{1}{4.5\pi\sigma} m^2 + ...
\label{eqn_adjointR}
\ee
where we have used the observed fact 
\cite{Deldar:1999vi}
that the adjoint and fundamental string tensions are related by
Casimir scaling, 
$\sigma_a \simeq \frac{9}{4} \sigma$ for $N=3$. This gives a slope
$\alpha^\prime \simeq 0.4 \, \mathrm{GeV}^{-2}$ which is very much flatter
than the usual mesonic Regge trajectory, although not quite
as flat as the phenomenological Pomeron or the leading glueball
trajectory we identified in Section~\ref{subsection_4D}.

Since the adjoint string comes back to itself under $C$, $P$ or
rotations of $\pi$, we expect its spectrum to contain
$J$ even and $P,C$ positive i.e.
\be
J^{PC} = 0^{++}, 2^{++}, 4^{++}, ...
\label{eqn_adjointQ}
\ee
just as one expects for an even-signature Pomeron.

\subsection{Closed strings}
\label{subsection_closed}
%
%
\begin{figure*}[t]
\vspace{-0.5cm}
\centerline{\begin{minipage}[c]{15cm}
\psfig{file=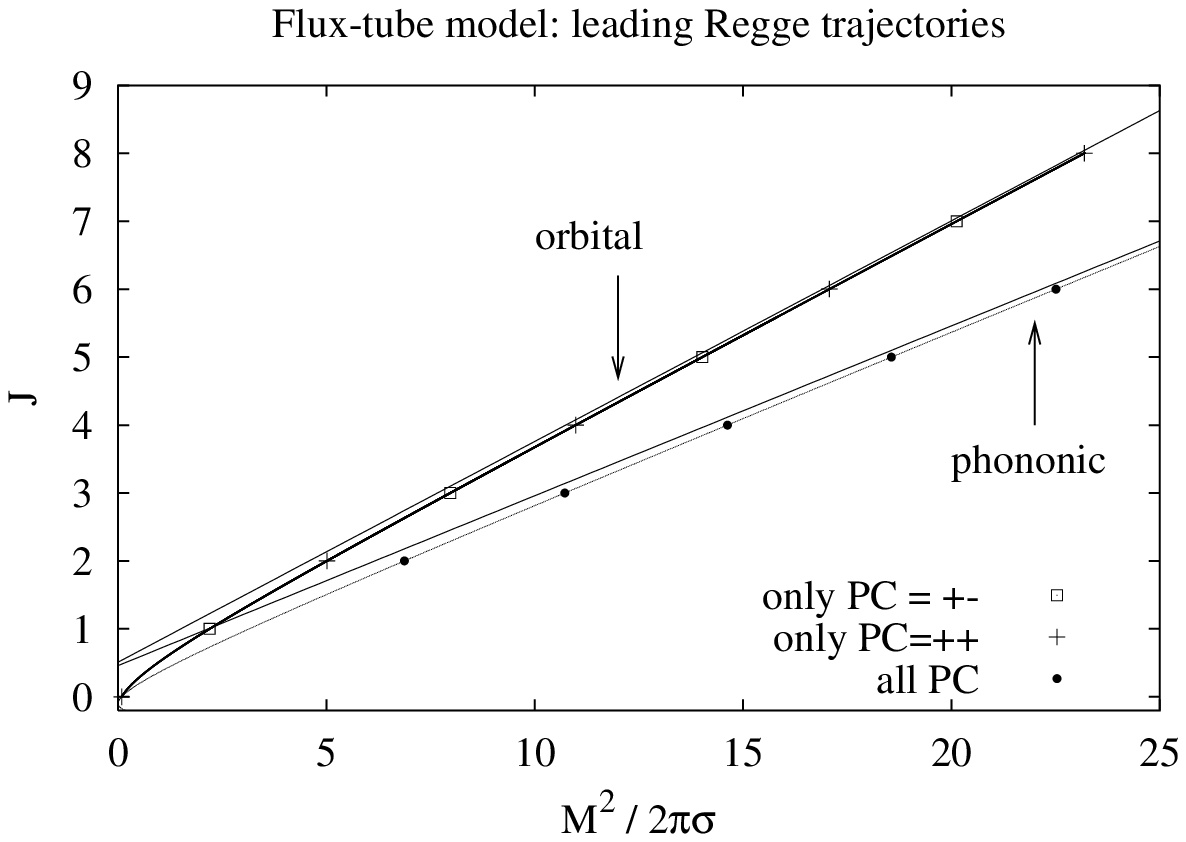,angle=0,width=15cm}
	    \end{minipage}}
\caption[a]{The leading phononic and orbital Regge trajectories in the 
flux-tube model in 3+1D (with the bosonic string Casimir energy correction).
The straight lines are semi-classical approximations to the trajectories. 
Crosses, circles and 
squares indicate the position of physical states with the corresponding
quantum numbers.}
\la{fig:traj}
\end{figure*}
%
For mesons an open string is the only natural string model.
For glueballs, however, an equally natural model is one composed
of a closed loop of fundamental flux with no constituent gluons
at all. This should not be regarded as an alternative 
model. Rather one expects some glueball states to be open strings
and others to be closed strings. (With mixing between the two, at
finite $N$.) Clearly we would like to identify which state
corresponds to which type of string.

An approximate but tractable closed string model was
constructed in
\cite{isgur}.
In this model the essential component is a circular
closed string (flux tube)  of radius $\rho$. There are 
phonon-like excitations of this closed string
which move around it clockwise or anticlockwise and
contribute to both its energy and its angular momentum.
The system is  (first) quantised so that we can calculate, 
from a Schr\"odinger-like wave equation
\cite{isgur}, 
the amplitude for finding a loop in a particular radius interval. 
The phonon excitations are regarded as `fast' so that they 
contribute to the potential energy term of the equation and
the phonon occupation number is a quantum number
labelling the wave-function. The whole loop can
rotate around its diameter, obtaining angular momentum
that way as well.

We refer the reader to
\cite{thesis}
for the details of our analysis of this model. Here we
simply state that if one considers the set of states where
the angular momentum is purely phononic one obtains an
asymptotically linear Regge trajectory with slope
\be
J = \alpha(t=m^2) 
\stackrel{J\to\infty}{=} \frac{1}{8\pi\sigma} m^2 + ...
\label{eqn_adjphon}
\ee
while for a loop with purely orbital motion one
obtains a linear trajectory with
\be
J = \alpha(t=m^2) 
\stackrel{J\to\infty}{=} \frac{3\surd 3}{32\pi\sigma} m^2 + ...
\label{eqn_adjorb}
\ee
In either case one obtains a slope 
$\alpha^\prime \simeq 0.2 - 0.3 \, \mathrm{GeV}^{-2}$ which is in
the right range for the Pomeron. One can also calculate
the intercept obtained by linearly extrapolating this trajectory
from large to small $J$ but this depends on both the string
`Casimir energy' correction and on any curvature term in the
effective string action. As an illustration we show in Fig.3
the Chew-Frautschi plot obtained by a numerical solution of the
model (with a conventional string Casimir energy and no
curvature term).

For the orbital trajectory, the geometry of the circular loop 
automatically gives it positive parity $P=+$. Furthermore, 
the mere fact that an oriented planar loop is spinning around an axis
contained in its plane implies that the charge conjugation is determined 
by the spin:
\be
P~=~+,\quad C~=~(-1)^{J},\quad J=0,1,2\dots
\ee
For the leading phononic trajectory, the most obvious feature is the 
absence of a $J=1$ state, because there is no corresponding phonon
(it amounts to a mere translation). Secondly, for a planar 
loop, parity has the same effect as a $\pi$-rotation around
an axis orthogonal to its plane. Therefore for phonons that
lie in the plane of the loop
\be
P~=~(-1)^J, \quad C=\pm
\quad J=0,2,3,4,\dots
\ee
while for phonons corresponding to fluctuations orthogonal
to that plane:
\be
P~=~(-1)^{J+1},\quad C=\pm
\quad J=0,2,3,4,\dots
\ee

\subsection{Other topologies}
\label{subsection_other}

It is conceivable that for those quantum numbers
for which the simple flux-tube model predicts a very large mass,
other topologies of the string provide ways
to construct a lighter fundamental state. A new pattern of quantum numbers 
arises 
if the oriented closed string adopts a twisted, `8' type configuration, 
whilst remaining planar. The parity of such an object is automatically locked to
the charge conjugation quantum number, $P=C$. The orbital trajectory
 built  on such a configuration leads to a sequence of states
\be
0^{++},~1^{--},~2^{++},~3^{--},~4^{++},~\dots
\ee
More exotic topologies of the string 
have been advocated in~\cite{Niemi:2003hb}, but they
presumably lead to more massive states. Such objects are more likely to be
relevant to the large $N$ limit, where they will not decay.

\subsection{String models in D=2+1}
\label{subsection_stringd3}

The SU(3) gauge theory in $D=2+1$ is linearly confining
and therefore an effective string theory description is equally well 
motivated. Since the rotating open string lies 
in a plane, it provides a natural model for glueballs in
two space dimensions. The closed string is also a possibility,
although now all the angular momentum must come from phonons
in the plane of the loop. 

The open adjoint string will contribute states with 
$J$ even and $C=+$, just as in \eqn(\ref{eqn_adjointQ})
except for the additional trivial parity doubling 
of non-zero spin states in two space-dimensions.

For the closed string, the quantum numbers for the leading $C=+$ 
and $C=-$ phononic trajectories are
\ba
J^{PC}&=&0^{++},~2^{P+},~3^{P+},~4^{P+},~\dots \quad
C=+\nonumber \\
J^{PC}&=&0^{--},~2^{P-},~3^{P-},~4^{P-},~\dots \quad
C=-,\nonumber
\ea
where $P$ is arbitrary, again because of the $J\not= 0$ parity doubling.
In the simplest form of the model, the two trajectories are degenerate.

We remark that an orbital trajectory could only be present if the string 
were to  acquire a `permanent deformation', as heavy nuclei can do,
but this goes beyond the scope of the simple flux tube model. 
The largest possible slope
is obtained in the extreme case of the collapse to a segment, 
when the slope is  $1/4\pi\sigma$. The twisted orbital trajectory
carries the states $0^{++}$, $1^{P-}$, $2^{P+}$, $3^{P-}$, \dots
($P$ arbitrary).

%
%
\section{Interpreting the glueball spectrum}
\label{section_interpret}
%
%

It is clear from the discussion in Section~\ref{section_strings}
that we need more than just the $PC=++$ spectrum if we
are to interpret  the observed glueball Regge trajectories
in terms of string models. We now present some results for
glueball states of other $P$ and $C$ and see how far we can interpret
the dynamics underlying the trajectories.

\subsection{Regge trajectories in D=3+1 }
\label{subsection_interpretd4}
%
%
\begin{figure*}[t]
\vspace{-0.5cm}
\centerline{\begin{minipage}[c]{14cm}
\psfig{file=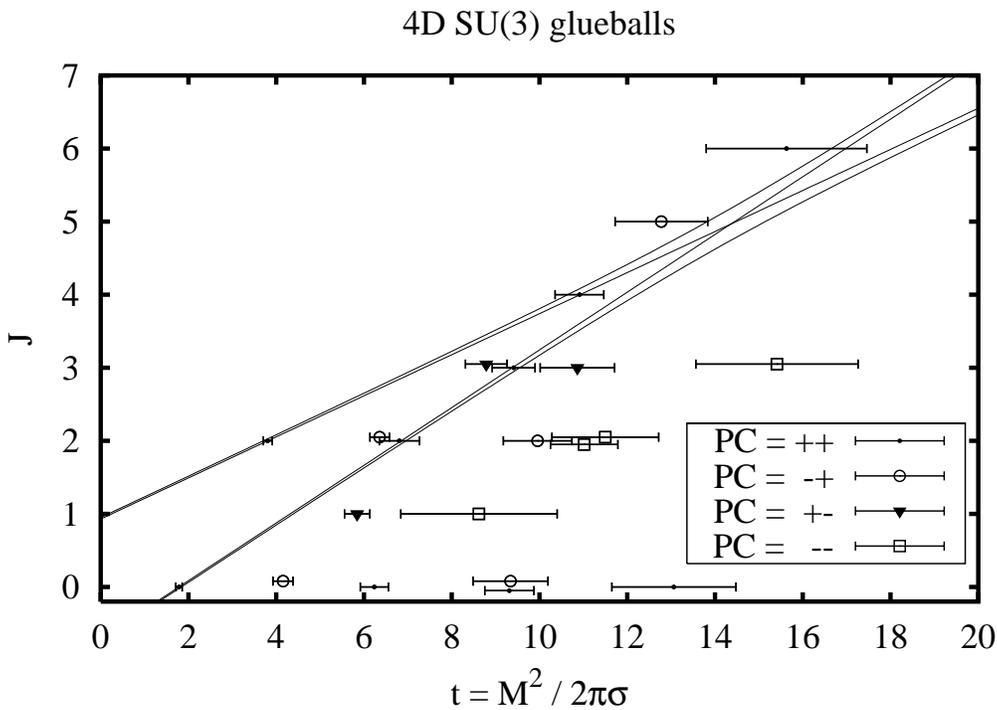,angle=0,width=14cm}	\end{minipage}}
\caption{Chew-Frautschi plot of the continuum 4D $SU(3)$ gauge theory.
}
\la{fig:cf}
\end{figure*}
%
\begin{figure*}[t]
\vspace{-0.5cm}
\centerline{\begin{minipage}[c]{14cm}
\psfig{file=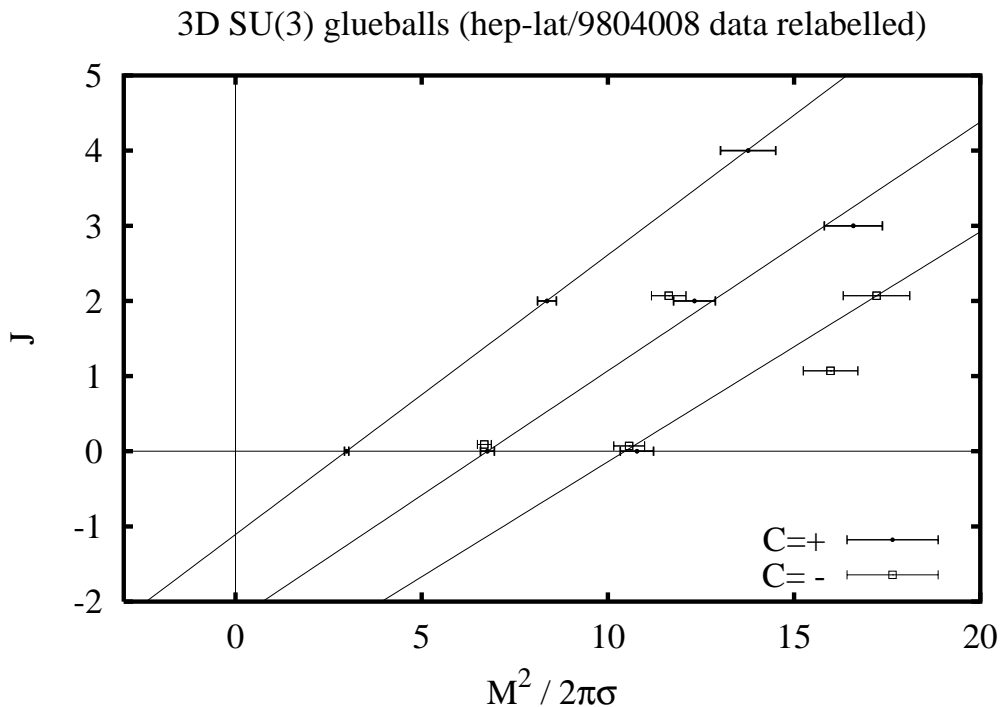,angle=0,width=14cm}	\end{minipage}}
\caption{The Chew-Frautschi plot of the continuum D=2+1 $SU(3)$ 
glueball spectrum.}
\la{fig:cf_su3_3d}
\end{figure*}

In Fig.4 we provide a Chew-Frautschi plot that contains not only
the $PC=++$ states already shown in Fig.1, but also the other states
that we have been able to identify in the continuum limit.

For $J\leq 4$ the leading trajectory contains only even spin
states with $PC=++$. This suggests that the trajectory arises
from a rotating open string carrying adjoint flux between the
gluons at the end points. 

The subleading trajectory has no $J=1$ state although it does
appear to have a $J=3$ and, possibly, a $J=5$ state. The absence
of the $J=1$ state (in the presence of other states of odd $J$)
is characteristic of the closed string phononic spectrum.
The parity doubling at $J=2$ (due to the near-degeneracy of
the lightest $2^{-+}$ and the first excited $2^{++}$) and
the near degeneracy of the lightest $3^{+-}$ and $3^{++}$
support this interpretation. We note that this is a non-trivial 
observation from the point of view of simple operator-dimension 
counting rules, since the  $3^{++}$ is created by a
dimension 5 operator and the $3^{+-}$ by a dimension 6 operator.
On the other hand, it seems that the expected light states with
quantum numbers $J~{\rm even}$, $C=-$ or $J~{\rm odd }$, $P=-$
are missing from the spectrum. It would be interesting 
to see whether string corrections to the flux-tube
model can provide a natural explanation for the corresponding
large mass splittings~\cite{thesis}.

Given that the two leading trajectories cross somewhere
around $J=5$ it is not clear to which trajectory we
should assign the observed $5^{-+}$ and $6^{++}$ states.
Our interpretation of the leading trajectory as being an open
string and the first sub-leading trajectory as being a phononic
closed string would require us to assign the  $5^{-+}$ to the 
latter and to expect an additional excited  $6^{++}$ close to
the ground state so that each trajectory would possess
a state with these quantum numbers.

We note that with the above interpretation, the open string 
trajectory has a smaller slope than that of the closed string in the 
small $J$ region. It is however plausible that at large $J$
(and in the absence of decays), the expected ratio of the slopes
(\eqn(\ref{eqn_adjointR}) and \eqn(\ref{eqn_adjphon})) would be restored.
Our interpretation could be tested by investigating the structure of the
fundamental and excited $2^{++}$ glueballs.

Looking to the heavier states, the fact that the $1^{--}$
is lighter than the $3^{--}$ is hard to understand within
the flux tube model. On the other hand it would arise naturally
from rotations of the `twisted' loop discussed in
Section~\ref{subsection_other}. (We refer to
\cite{thesis}
for a detailed discussion of this possibility.)

\subsection{Regge trajectories in D=2+1 }
\label{subsection_interpretd3}

In Fig.5 we present a Chew-Frautschi plot for the SU(3)
gauge theory in 2 spatial dimensions, with both the
$C=+$ and $C=-$ states displayed (using a relabelling of
the states in
\cite{teper98}).
We suppress the $P$
label because of the automatic parity-doubling for
$J\not= 0$ states.

The leading trajectory contains only even $J$ states
with $C=+$ and so is naturally interpreted as arising
from a rotating open (adjoint) string. Since the
intercept is sufficiently  low, it can and does include a 
$J=0$ state, in contrast to the case of 3 spatial dimensions.

The first subleading trajectory has no $J=1$ state, although
it contains a $J=3$ state, and possesses a $C=+/-$ degeneracy
for the lower $J$ where we have reliable calculations.
All this strongly suggests a closed string interpretation.
This will necessarily be phononic, since there are no
non-trivial global rotations  of a circular loop in two
space dimensions.

\subsection{The odderon}
\label{subsection_odderon}

There is some experimental evidence, from the difference
between $pp$ and $p{\bar p}$ differential cross-sections
at larger $t$, for an odd signature $C=-$ trajectory that
is $very$ flat, $\alpha^\prime \sim 0$, and that has 
a (near) unit intercept, $\alpha(0) \simeq 1$. This has
been named the `odderon'
\cite{odderon}.

The states one might expect to lie along the odderon are the
lightest $1^{--}$, $3^{--}$, $5^{--}$, ... glueballs. From Fig.4
we see that a trajectory defined by the lightest $1^{--}$ and 
$3^{--}$ will have a slope similar to the Pomeron and a very low,
negative intercept. (Such a trajectory also passes through the lightest 
$2^{--}$, suggesting an exchange degenerate trajectory of opposite
signature.) From this point of view, our spectrum provides no evidence 
in favour of the phenomenological odderon being the leading $PC=--$
glueball trajectory.

However there is a (significant) caveat. If the leading trajectory 
has an intercept around unity, as claimed phenomenologically,
then the lightest $1^{--}$ glueball cannot lie on it, but will
rather lie on a subleading trajectory. To test this possibility
we need a good calculation of the lightest $5^{--}$ glueball,
something we do not have at present. We finish by noting that if 
we simply draw a linear trajectory from $J=1$ through the mass
of the lightest $3^{--}$ glueball, we obtain an `odderon' slope
that is about half the Pomeron slope, which is in the direction
of the phenomenological expectation.

%
%
\section{Large N}
\label{section_largeN}
%
%
One does not expect the leading glueball trajectory
to be exactly like the `Pomeron' both  because the higher-$J$ 
states are unstable and because in the real world 
there will be mixing between glueballs and flavour-singlet 
$q\bar{q}$ mesons. It is only in the limit of SU$(N\to\infty)$
that one might expect Regge trajectories to be exactly linear 
(no decays) and the leading glueball Regge trajectory to be 
precisely the Pomeron (no mixing)~\cite{regge_2d}.

It is therefore interesting to ask if the SU(3) glueball spectrum 
is close to that of  SU$(N\to\infty)$. Although recent lattice 
calculations
\cite{lucini,lucini04}
have demonstrated that this is so for the lightest $0^{++}$ and 
$2^{++}$ glueballs, that is too limited a result for our purposes.
We have therefore computed the glueball spectrum in SU(8) by 
similar techniques to those we have used in SU(3). Since the leading
large-$N$ correction is expected to be $O(1/N^2)$, we can
be confident
(see also
\cite{lucini,lucini04})
that $N=8$ will be very close to $N=\infty$ for most physical 
quantities. Leaving the details of this calculation to
\cite{thesis}
we simply compare in Fig.6 the low-lying 
SU(3) and SU(8) continuum glueball
spectra. We see a close similarity except for the first excited
$0^{++}$ (upon which we will comment below). Although  the
accuracy of this calculation did not permit us to identify
higher-spin glueballs, we take this as evidence that the
leading glueball trajectories at $N=3$ and $N=\infty$
will be very similar.

While the low-lying spectrum may not change much when we go from 
$N=3$ to $N=\infty$, the string picture suggests an interesting 
way in which the excited state spectrum may alter as $N$ increases.
This arises because there are more stable flux tubes than just
the fundamental one at larger $N$ (see e.g.
\cite{Lucini:2001nv}).
These are called $k$-strings, they have string tensions 
$\sigma_k<k\sigma$, and the number of distinct strings is equal to 
the integer part of $N/2$. Thus there should be a separate sector of the 
glueball spectrum based on closed loops of each of these $k$-strings.
These sectors will be identical except that they will be rescaled
by $\sqrt{\sigma_k/\sigma}$. This is a striking prediction.
In particular, since we have identified the lightest $0^{++}$
as being a closed string of fundamental ($k=1$) flux, we would
expect the lightest  $0^{++}$ based on the $k=2$ closed string
to have a mass $m^\star_{0^{++}} \simeq  1.34 m_{0^{++}}$
taking the value of  $\surd\sigma_{k=2}/\surd\sigma$ for $N=8$
from 
\cite{lucini04}.
It is interesting to note that the anomalously light excited scalar
that we observed in SU(8) fits this expectation quite well. It may
constitute the first observation of one of these new states.

We remark that other, unstable strings which become
stable as $N\to\infty$ may have further implications for the
glueball spectrum at larger $N$.

\begin{figure*}[t]
\centerline{\begin{minipage}[c]{17cm}
   \psfig{file=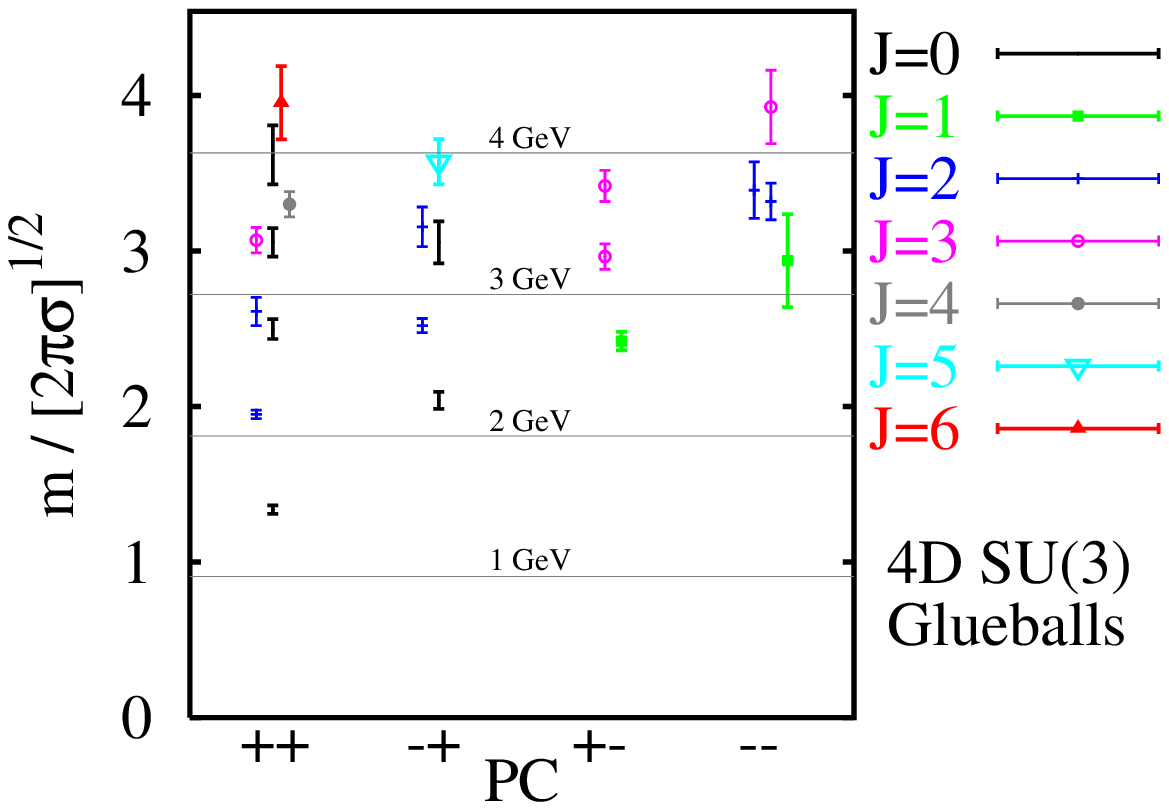,angle=0,width=8.5cm}
   \psfig{file=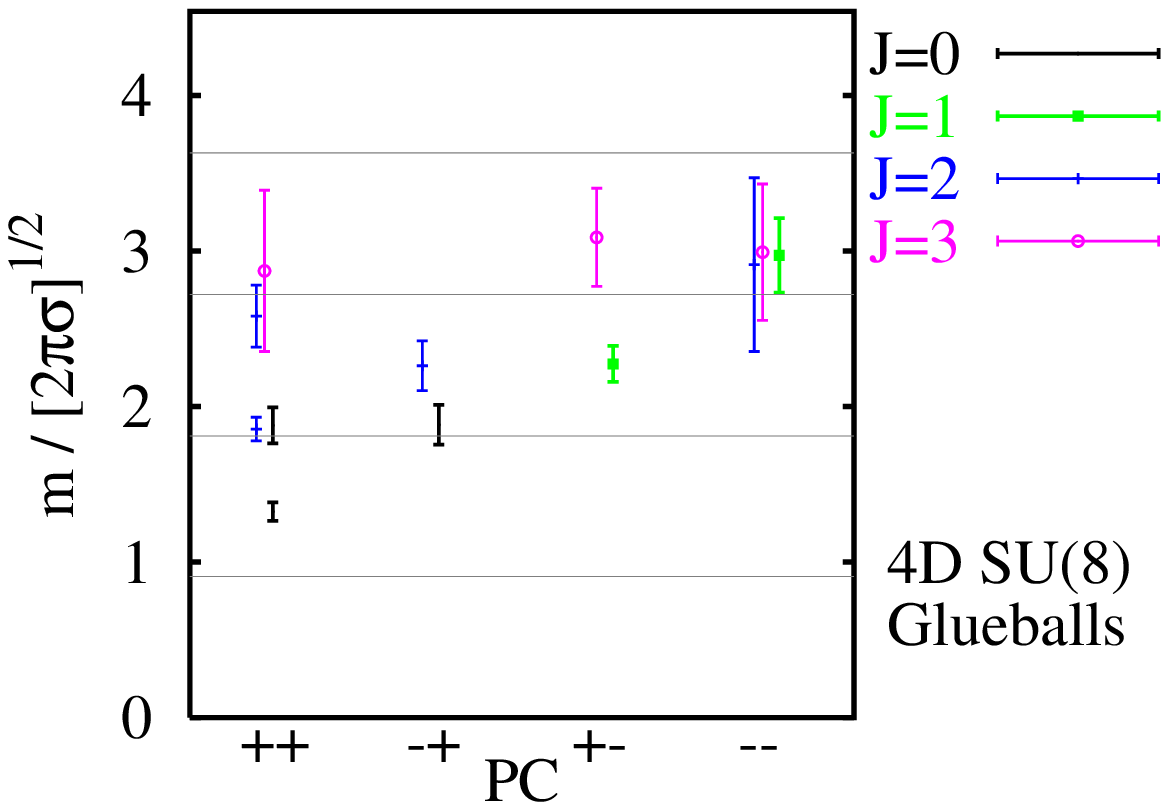,angle=0,width=8.5cm}
	    \end{minipage}}
\caption{The continuum spectrum of glueballs in the 4D pure $SU(3)$ and 
$SU(8)$ gauge theories. The physical scale was set using 
$\sqrt{\sigma}=440$MeV.}
\la{fig:spec_all}
\end{figure*}

%
%
\section{Conclusions}
\label{section_conclude}
%
%

Using novel lattice techniques, we have calculated the masses of 
higher spin glueballs in the continuum limit of the SU(3) gauge
theory. In the physically interesting case of 3+1 dimensions 
we find a leading $PC=++$ glueball trajectory 
\be
\alpha_P(t) = 0.93(24) + 0.25(2) t
\label{eqn_pomeronG}
\ee
(in GeV units, using a conventional value of the string tension, 
$\surd\sigma \simeq 420 \, \mathrm{MeV}$, and assuming linearity) 
which is entirely consistent with the phenomenological Pomeron. 
The sub-leading trajectory has a larger slope and eventually 
`crosses' the Pomeron. We argue that such a rich Regge structure 
for glueballs occurs naturally within string models: while quarkonia
arise only from open strings (of fundamental flux joining two quarks),
glueballs can arise not only from open strings (of adjoint
flux, joining two gluons), but also from closed strings (closed
loops of fundamental flux). The latter obtains its angular momentum
both from two kinds of `phonons' running around the perimeter 
and from rotations of the whole loop around a diameter.
Asymptotic calculations suggest an interesting structure of 
non-parallel as well as parallel trajectories. Whether this
might bear upon the existence of the `hard' Pomeron is
an interesting question. 

To try and identify the dynamical content of the different
trajctories, we also calculated states with other $P$ and $C$.
We then argued that the states on the Pomeron are given by a rotating
open string while the sub-leading trajectory has the characteristics 
of a closed string whose spin comes from phonons running around
in the plane of the loop.

In contrast to this, we find that in 2+1 dimensions the intercept
of the leading trajectory is negative so that the Pomeron in
that case does not contribute significantly to scattering at 
high energies. Here again we find evidence that the leading trajectory
is an open string while the non-leading one is a closed string.
In this case we have enough accurately calculated glueball states 
along the leading trajectory to demonstrate its approximate linearity.

Of course it is only at $N=\infty$ that one can expect Regge 
trajectories to be exactly linear
and glueballs to define the physical Pomeron.
We showed through a calculation of the SU(8) glueball spectrum 
that SU(3) is indeed close to  $N=\infty$ for the low-lying
glueball spectrum with a single striking exception that we interpreted
as the first signal of the new closed $k$-string states    
one expects to appear at higher $N$.

Finally, we briefly comment upon high energy scattering.
As $N\to\infty$ the usual counting arguments tell us that
scattering amplitudes vanish. So at large $N$ we expect
the partial waves to be far from the unitarity limit, i.e.
little shadowing, and so the additive quark counting rule
for Pomeron coupling to hadrons is natural. 
The experimentally observed additive quark rule thus 
constitutes one more indication that QCD is `close' to SU($\infty$).
If the Pomeron intercept is higher than 
unity, then at high enough energy this will break down,
and shadowing will become important so that the cross-section
can satisfy the Froissart bound.

In a world with only bottom quarks,
the Froissart bound $\sigma_{\rm tot}\leq \frac{\pi}{m_G^2}
\log^2\left(\frac{s}{s_0}\right)$ is stronger by two orders of magnitude.
Our glueball data strongly suggests that high-energy cross-sections are
approximately constant  in the quenched world and that its 'pomeron'
trajectory has properties very similar to the real-world pomeron.
It provides a (partial) justification for perturbative analyses 
that are based on the gluon field only and are meant to
describe the real world. But it is clear that in such frameworks,
unitarisation should be enforced with respect to the gluonic Froissart bound.

We can also turn the argument around.  Experimentally, the
high-energy $pp$ cross-section lies only slightly  
under the Froissart bound of gluodynamics for $m_G\simeq1.6$GeV.
If the $pp$ cross-section is found to exceed it at the Large Hadron Collider,
then it will definitely
be necessary to include the effects of light quarks in the description
of the hadronic wave-functions at that energy.
%
%
\section*{Acknowledgements}
%
%

The numerical calculations were performed on a PPARC and EPSRC
funded  Beowulf cluster in Oxford Theoretical Physics.
H.M. thanks  the Berrow Trust for financial support.


\end{document}